\documentclass{aastex631}

\begin{document}

\title{Formation of Near-surface Atmospheric Inversion and Surface Inversion in Hothouse Climates}

\correspondingauthor{Feng Ding, Jun Yang}
\email{fengding@pku.edu.cn, junyang@pku.edu.cn}

\author[0000-0001-8981-1798]{Jiachen Liu}
\affiliation{Laboratory for Climate and Ocean-Atmosphere Studies, Department of Atmospheric and Oceanic Sciences, School of Physics, Peking University, Beijing 100871, China}

\author[0000-0001-7758-4110]{Feng Ding}
\affiliation{Laboratory for Climate and Ocean-Atmosphere Studies, Department of Atmospheric and Oceanic Sciences, School of Physics, Peking University, Beijing 100871, China}

\author[0000-0001-6031-2485]{Jun Yang}
\affiliation{Laboratory for Climate and Ocean-Atmosphere Studies, Department of Atmospheric and Oceanic Sciences, School of Physics, Peking University, Beijing 100871, China}

\begin{abstract}
A hothouse climate may develop throughout Earth's history and its warming future and on potentially habitable exoplanets near the inner edge of the habitable zone. Previous studies suggested that near-surface atmospheric inversion (NAIV) with planetary boundary air temperature being higher than the air temperature adjacent to the surface, is a pronounced phenomenon in hothouse climates. However, the underlying mechanisms are unclear. Here we show that lower-tropospheric radiative heating is necessary but not independently sufficient in forming the NAIV. Instead, the dynamic heating induced by large-scale subsidence is essential. With the prescribed reasonable large-scale subsidence, NAIV appears in small-domain cloud-resolving simulations, which was not observed in previous studies. Surface evaporative cooling also contributes to the formation of the NAIV. Besides NAIV, we find that surface inversion (SIV) with the air adjacent to the surface being warmer than the underlying sea surface is also a distinct phenomenon in hothouse climates. SIV is caused by strong surface evaporative cooling and large atmospheric shortwave absorption. These two types of inversion strongly stabilize the atmosphere, weaken atmospheric circulation, dry the free troposphere, and suppress the hydrological cycle. 
\end{abstract}


\section{Introduction} \label{sec:intro}
Within Earth's long history and future, the surface temperature (T$_s$) varies over a wider range than the present-day climate. Geological records suggest that Earth may have experienced an extremely hot climate (T$_s> \sim$70$^\circ C$) in the early Archean \citep{knauth2003high,robert2006palaeotemperature,sleep2010hadean}. Combined with observations of glacial deposits and other evidence, it is generally believed that Earth has fallen into a snowball state in the Neoproterozoic era \citep{hoffman1998neoproterozoic}. Simulations show that a very high CO$_2$ concentration ($\sim$100,000 ppmv or higher) is needed to deglaciate a snowball Earth \citep{pierrehumbert2005climate,LEHIR2007274,Hu}. The strong greenhouse effect induced by the high concentrations of CO$_2$ and H$_2$O can trigger a hothouse climate (T$_s> \sim$50$^\circ C$) after deglaciation \citep{le2009snowball,yang2017persistence,hoffman2017snowball}. A high concentration of CO$_2$ may also lead to hothouse climates in other epochs of Earth's history or the distant future \citep{wordsworth2013water,popp2016transition,wolf2018evaluating}.
In addition, it is predicted that the Sun has a gradually brightening fate as a main-sequence star, so our Earth will inevitably receive more solar flux and undergo a hothouse climate \citep{gough1981solar,ribas2009sun,Leconte_Forget_Charnay_Wordsworth_Pottier_2013, wolf2015evolution}.
Potentially habitable exoplanets near the inner edge of the habitable zone may also experience a hothouse climate. Exploring the features of hothouse climates helps us understand the climate evolution of early Earth, predict the climate of future Earth, and examine the habitability of exoplanets near the inner edge of the habitable zone.

A pronounced phenomenon in the hothouse climate simulations is that a near-surface atmospheric inversion (NAIV, within which the air adjacent to the surface is cooler than the air above it) develops, making the atmospheric stratification, convection, large-scale circulation, and hydrological cycles very different from modern Earth's climate. Using a one-dimensional (1D) radiative-convective model, \citet{wordsworth2013water} found that a strong NAIV forms when T$_s$ is fixed at 350 K. \citet{wolf2015evolution} observed a permanent planetary-scale NAIV in mid-latitudes when T$_s$ reaches $\sim$330 K in general circulation model (GCM) simulations. This NAIV grows stronger and deeper as T$_s$ increases. \citet{popp2016transition} also observed an inversion near the surface as T$_s$ climbed to $\sim$330 K. However, the underlying mechanisms of NAIV in the hothouse climate have not been adequately investigated. 

Besides occurring in hothouse climates, NAIV is also observed under Earth's present climate, where its mechanism has been thoroughly studied. Previous studies \citep[e.g.,][]{schubert1995dynamical,johnson1999trimodal,carrillo2016characterization,https://doi.org/10.1002/joc.7151} have documented and analyzed the subtropical NAIV, commonly known as the trade-wind inversion. This inversion is primarily driven by large-scale subsidence associated with the descending branch of the Hadley cell and the upwelling of cold ocean water. However, compared to NAIV in hothouse conditions, the present-day subtropical NAIV is confined to limited regions over the eastern subtropical oceans, where the ocean heat uptake is stronger. 
Temperature inversions also frequently occur within the polar boundary layer under the current climate, mainly driven by strong surface radiative cooling and the advection of warm air aloft \citep[e.g.,][]{Bradley_Keimig_Diaz_1992,Bradley_Keimig_Diaz_1993}. Given the substantial climatic differences between hothouse and modern conditions \citep[e.g.,][]{wolf2015evolution,popp2016transition,seeley2021episodic}, the formation mechanisms of NAIV in hothouse climates may also differ significantly from those in the present day.

One hypothesis is that NAIV in the hothouse climate is induced by lower-tropospheric radiative heating (LTRH) \citep{wordsworth2013water,wolf2015evolution}. LTRH is caused by the closing of the atmospheric window by the absorption of water vapor in the thermal infrared. In moderate climates like modern Earth's, the absorption of water vapor between 8 and 13 $\mu$m is weak, so longwave radiation from the surface and the lower atmosphere can be emitted to space through this atmospheric window region \citep{goody1995atmospheric,pierrehumbert2010principles,koll2018earth,seeley2021h2o}. Regardless of the forcing mechanism, when the climate is hot and very humid, the continuum absorption of water vapor increases in the window region dramatically, hindering the longwave emission from going out, resulting in a weak infrared radiative cooling \citep{pierrehumbert2010principles}. Therefore, the absorption of near-infrared solar radiation by water vapor induces net radiative heating in the lower atmosphere \citep{wordsworth2013water}. However, recently, in the hothouse climate simulations in \citet{seeley2021episodic,seeley2023moist} using a small-domain cloud-resolving model (CRM), LTRH is observed, but NAIV does not appear, questioning the hypothesis.

In this study, we aim to investigate the underlying mechanisms of the formation of NAIV and to explain why it appears in GCM simulations but not in CRM. In addition, we observe the occurrence of surface inversion (SIV) in our climate simulations, another distinct phenomenon found in hothouse climates, and explain its formation. 

The outline of the paper is as follows. In Section \ref{methods}, we introduce the GCMs and CRM we used in this paper and our experimental designs. In Section \ref{phenomena}, we describe the features of NAIV and SIV in hothouse climates and explain their formation mechanisms in Section \ref{mechanism}. To further illustrate the formation mechanism of NAIV and SIV, we develop a 1D conceptual model in Section \ref{1D model}. Finally, conclusions and discussion are given in Section \ref{conclusions}.

\section{Model Descriptions and Experimental Designs}\label{methods}

We employ two GCMs to simulate the global hothouse climate, Community Earth System Model (CESM) version 1.2 and ExoCAM. We also use one CRM to explore the hothouse climate in small-domain simulations.

\begin{figure}
\centering
\includegraphics[width=\linewidth]{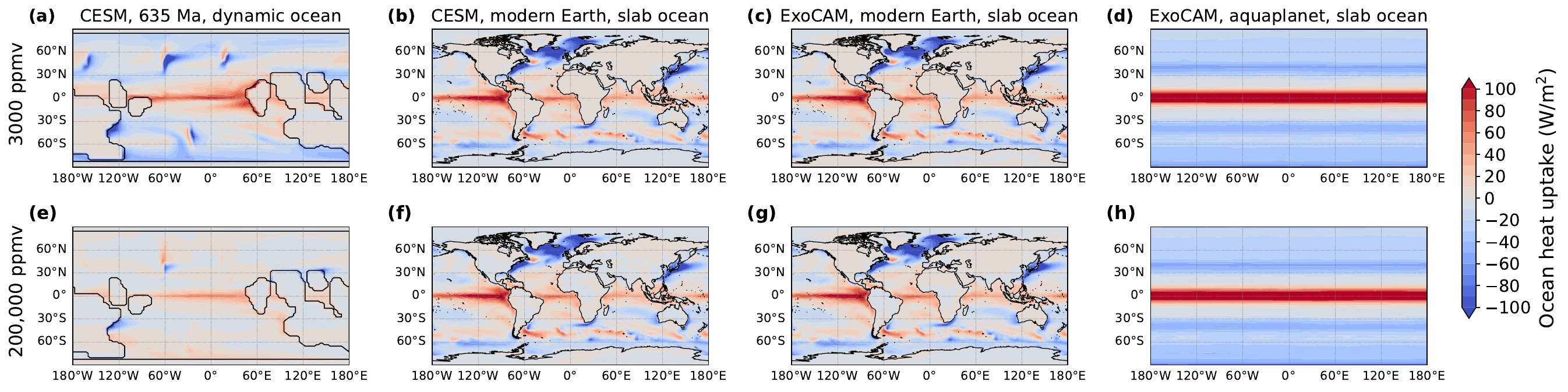}
\caption{The horizontal distribution of ocean heat uptake. Positive values indicate that the ocean gains energy from the atmosphere, while negative values indicate that the atmosphere gains energy from the ocean. The first column presents simulation results from experiments conducted with CESM using the 635 Ma paleogeography and a dynamic ocean. The second, third, and fourth columns depict the prescribed ocean heat flux used in slab ocean experiments conducted with CESM using modern Earth's land configuration, ExoCAM using modern Earth's land configuration, and ExoCAM aquaplanet simulations, respectively.}
\label{Figure_ocean_heat}
\end{figure}

CESM1.2.1 is a fully coupled climate model consisting of ocean, atmosphere, land, and sea ice components \citep{TheCommunityEarthSystemModelAFrameworkforCollaborativeResearch}. Two different land configurations are used in the simulations. One is similar to modern Earth's geography, but replacing the permanent land glaciers in Greenland, Antarctica, and the Himalayas with bare soil. The other is the reconstructed global paleogeography 635 million years ago \citep{LI2008179}, with land all covered with loam, an average soil in CESM \citep{rosenbloom2011using}. For simulations of modern Earth's geography, we use an atmosphere-land coupled model with a finite volume grid at a resolution of 1.9$^{\circ} \times$ 2.5$^{\circ}$ (latitude by longitude). The ocean component is a 50-m slab ocean, with prescribed ocean heat transport (Figure \ref{Figure_ocean_heat}(b)) similar to modern Earth \citep{bitz2012climate}.  The simulations use 26 vertical atmospheric levels. The eccentricity, obliquity, and rotation periods are identical to modern Earth. We use solar spectrum in the simulations. The insolation is set to 1290 W m$^{-2}$ as that in 635 Ma, 6\% lower than present. The atmosphere is assumed to consist of 0.78 bar N$_2$ with variable CO$_2$ and H$_2$O. The concentration of CO$_2$ is set to 3000 ppmv and 200,000 ppmv, with the first inducing a moderate climate, and the latter resulting in a hothouse climate after the melting of a hard-snowball Earth. These slab ocean experiments are run for 40 to 50 years and have reached equilibrium. The average of the last 10 years is used in the analysis.

For the 635 Ma configuration, a fully coupled set is used, with a finite volume grid at a resolution of 1.9$^{\circ} \times$ 2.5$^{\circ}$ (latitude by longitude) for the atmosphere and land components, and an approximate 1$^{\circ}$ displaced pole grid for the ocean and ice components. The dynamic ocean depth is 3500 m. The atmospheric composition, stellar flux, orbital parameters, and atmospheric vertical resolution are identical to the modern Earth's geography simulations.
The fully coupled simulations have run for over 1000 years with energy imbalance within 2 W m$^{-2}$ at the model top, and the average of the last 200 years is used in the analysis.  

ExoCAM\footnote{ https://github.com/storyofthewolf/ExoCAM} utilizes the Community Atmosphere Model version 4 \citep{neale2013mean,wolf_exocam_2022}. We use a 1.9$^{\circ}$ $\times$ 2.5$^{\circ}$ (latitude versus longitude) horizontal resolution and 26 vertical layers. ExoCAM uses a correlated-k two-stream radiative transfer model ExoRT\footnote{ https://github.com/storyofthewolf/ExoRT}, which is modified to simulate hothouse climates and high-CO$_2$ atmospheres \citep{wolf2013hospitable}. ExoCAM simulations are conducted with both modern Earth's land configuration while replacing permanent land glaciers with bare rock and an aquaplanet (represents exoplanets covered by global water oceans). The atmosphere is comprised of 0.78 bar N$_2$ with variable CO$_2$ and H$_2$O. The CO$_2$ concentration is set to 3000 ppmv and 200,000 ppmv as simulations using CESM. The solar spectrum is used in the simulations. The insolation is 1290 W m$^{-2}$ and the orbital-rotation parameters are identical to present-day Earth. The atmosphere is coupled to a 50-m slab ocean. In experiments with Earth's geography, we use a prescribed ocean heat transport (Figure \ref{Figure_ocean_heat}(c)) similar to present-day Earth \citep{bitz2012climate}. In aquaplanet experiments, ocean heat transport is prescribed identically to a previously done fully-coupled aquaplanet simulation (Figure \ref{Figure_ocean_heat}(d)).
The experiments are run for 40 to 60 years to reach equilibrium and the average of the last 10 years is used in the analysis.

The CRM employed in this study is the System for Atmospheric Modeling (SAM) version 6.11.6 documented by \citet{khairoutdinov2003cloud}. 
In this study, we use a single-moment microphysics scheme \citep{khairoutdinov2006high} and the radiative transfer model adapted from the Community Atmospheric Model version 3.0 \citep{collins2004description,collins2006formulation}.
All SAM simulations are conducted on doubly periodic 72 $\times$ 72 km$^2$ small domains with 144 vertical levels extending to 64 km. The horizontal resolution in both directions is 2 km. The vertical grid spacing is 25 m below 650 m, 500 m between 5.4 and 33 km, and 1 km above 38 km. The atmosphere consists of 0.8 bar N$_2$, 0.2 bar O$_2$, 400 ppmv CO$_2$, and variable H$_2$O. The solar spectrum is used in the simulations. The surface temperature is fixed at 325 K, and no planetary rotation and diurnal cycle are included. These experiment designs are similar to \citet{seeley2021episodic} to provide a better comparison with their simulated results.

\section{Results}\label{result}
\subsection{Near-surface Atmospheric Inversion and Surface Inversion in Hothouse Climates}\label{phenomena}

\begin{figure}
\noindent\includegraphics[width=\textwidth]{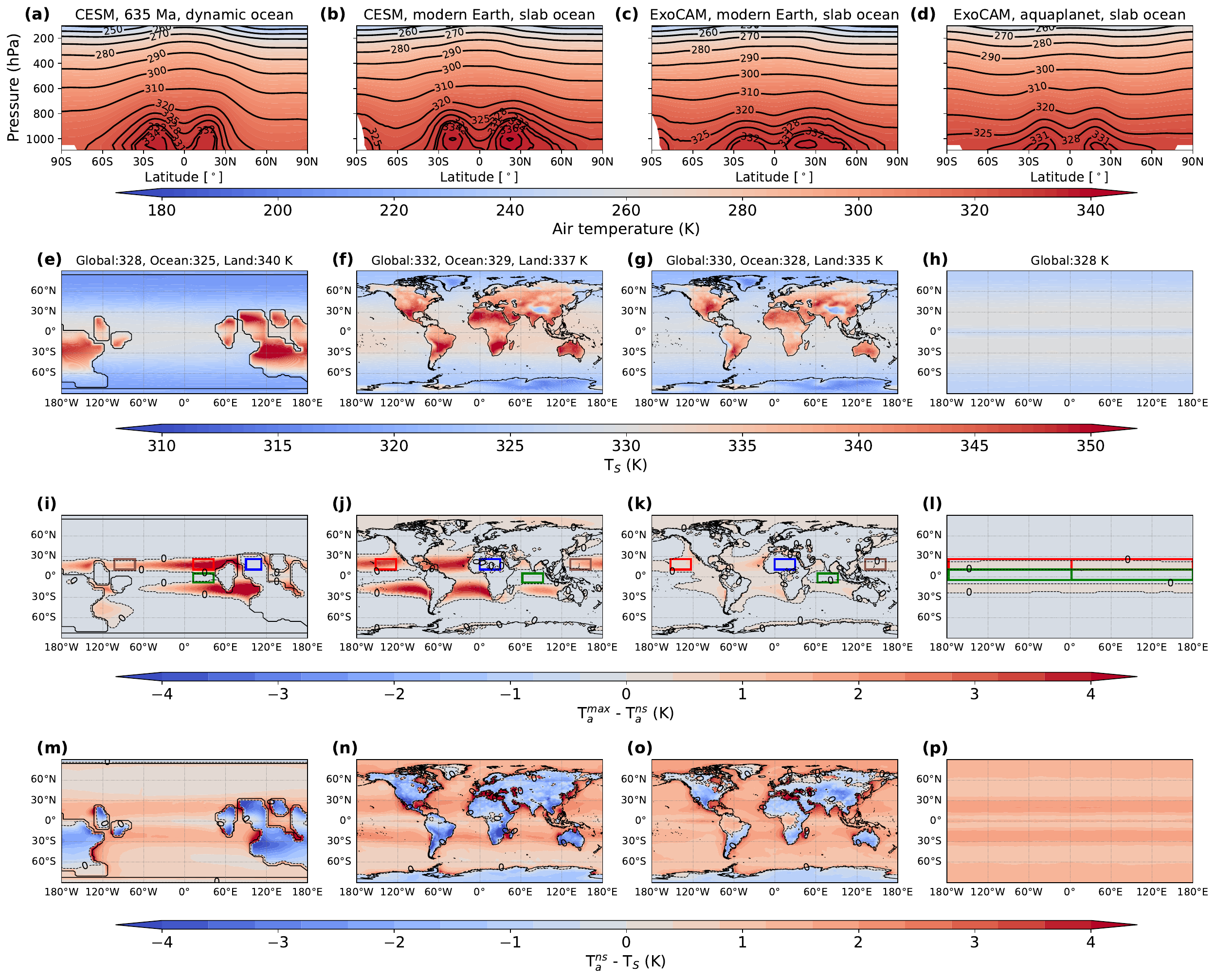}
\caption{Near-surface atmospheric inversion and surface inversion in GCM simulations. Panels (a\textendash d) show the latitude versus pressure distribution of air temperature. Panels (e\textendash h) show the horizontal distribution of surface temperature. The global-, ocean-, and land-mean surface temperatures are printed above. Panels (i\textendash l) show the horizontal distribution of the near-surface atmospheric inversion (defined as $T_{a}^{max} - T_{a}^{ns}>0$). Panels (m\textendash p) depict the horizontal distribution of the surface inversion (defined as $T_{a}^{ns} - T_s>0$).  Colored rectangles in panels (i\textendash l) highlight specific areas for further analysis: red and brown rectangles indicate descending areas over the ocean, green rectangles represent ascending areas over the ocean, and blue rectangles signify ascending areas over the land. 
Columns from left to right are simulation results from experiments with 635 Ma paleogeography using CESM, modern Earth's land configuration using CESM, modern Earth's land configuration using ExoCAM, and aquaplanet using ExoCAM, all with 200,000 ppmv CO$_2$.
\label{Figure_200000ppmv}}
\end{figure}

\begin{figure}
\noindent\includegraphics[width=\textwidth]{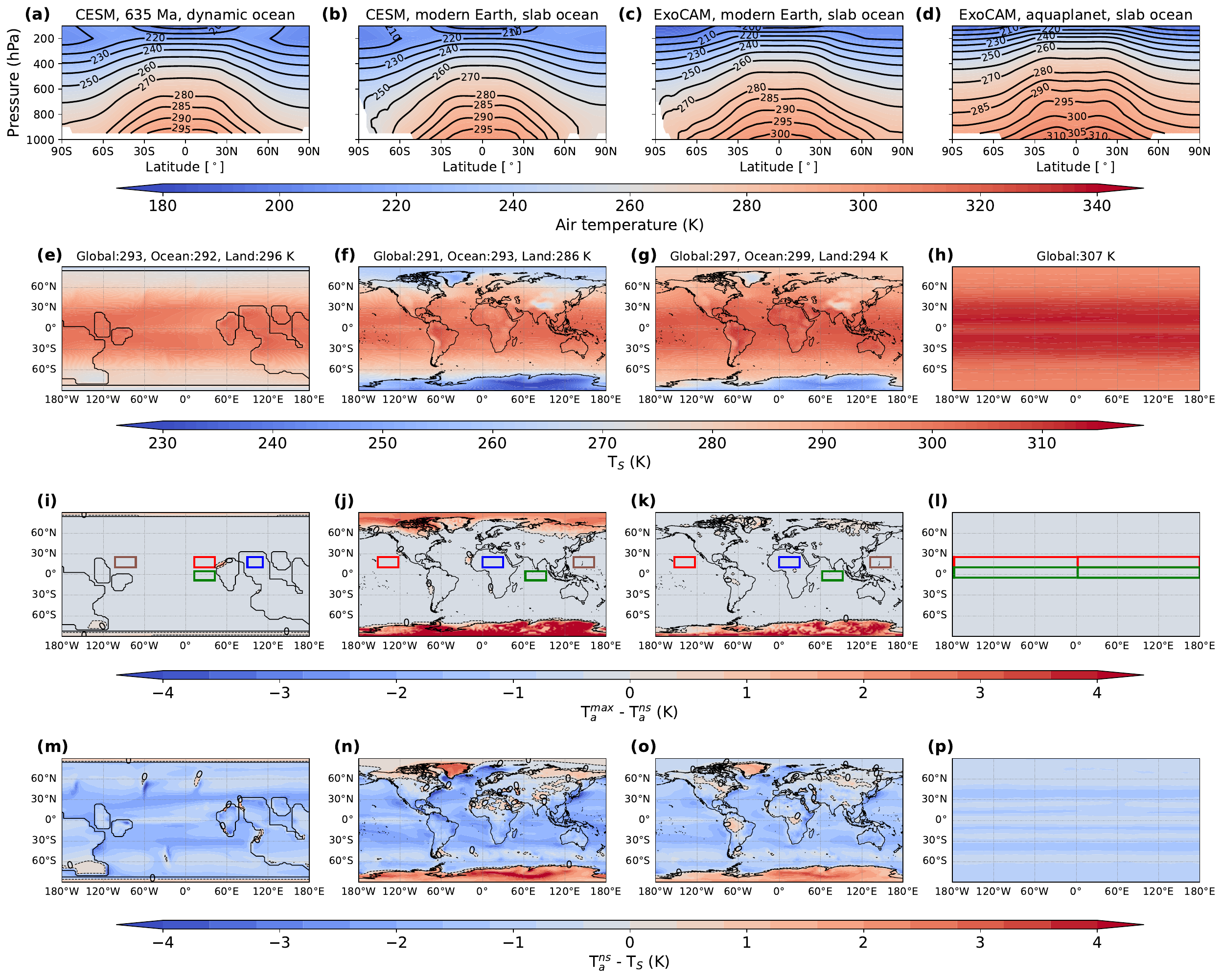}
\caption{Same as Figure \ref{Figure_200000ppmv}, but for experiments with 3000 ppmv CO$_2$.}
\label{Figure_3000ppmv}
\end{figure}

\begin{figure}
\centering
\includegraphics[width=\textwidth]{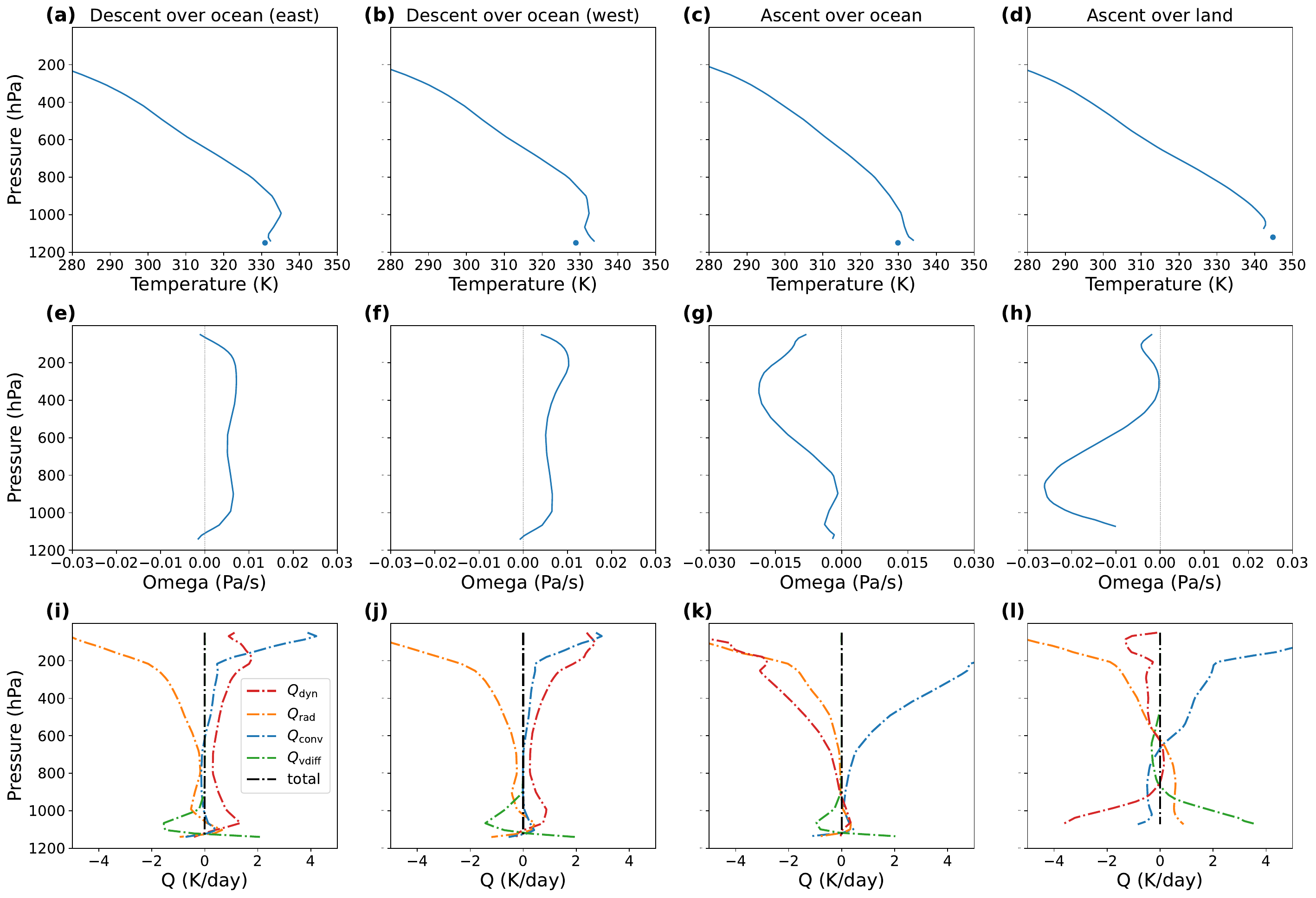}
\caption{Profiles of air temperature (a\textendash d), vertical velocity (e\textendash h), and temperature tendencies (i\textendash l).
The results presented from left to right are the regional averages within the areas enclosed by the red, brown, green, and blue rectangles in Figure \ref{Figure_200000ppmv}(j). Blue dots in panels (a\textendash d) show the surface temperature of each area. In panels (e\textendash h), positive values indicate subsidence and negative values indicate ascent. Dash-dotted lines in panels (i\textendash l) represent temperature tendencies induced by dynamic heating (red), radiation (orange), convection (blue), vertical diffusion (green), and the total temperature tendency (black).
Results are from the CESM experiment with modern Earth geography and 200,000 ppmv CO$_2$.
\label{Figure_profile}}
\end{figure}

\begin{figure}
\centering
\noindent\includegraphics[width=\textwidth]{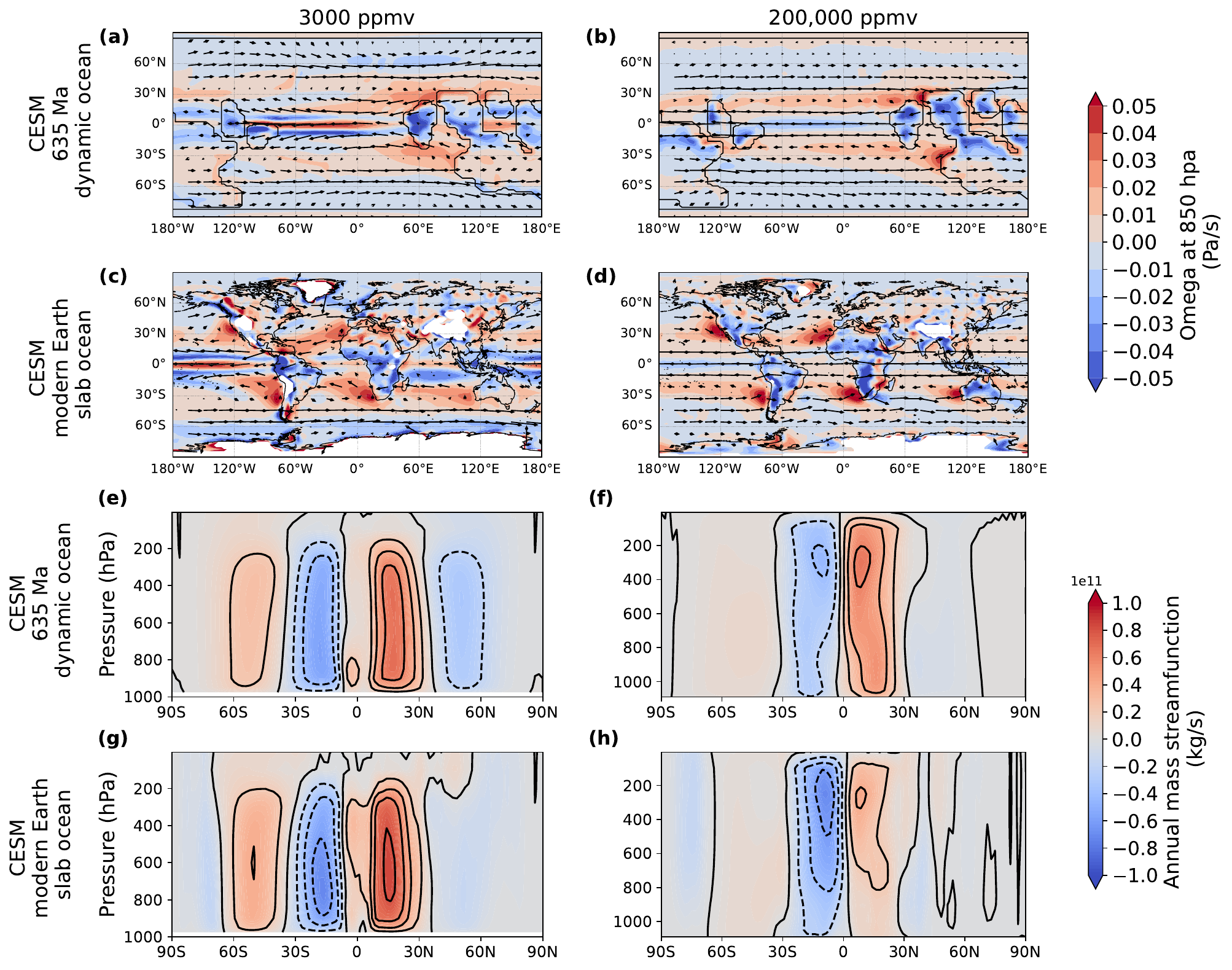}
\caption{Horizontal distribution of pressure velocity at 850 hPa (a-d) and latitude versus pressure distribution of annual mean mass stream function (e-h).
In panels (a-d), contours represent the horizontal distribution of pressure velocity at 850 hPa and vectors illustrate the horizontal wind at the same level. Positive pressure velocity values indicate subsidence, while negative values correspond to ascent. In panels (e-h), the lower and upper bounds are -10$^{11}$ and 10$^{11}$ kg s$^{-1}$ with intervals of 2$\times$10$^{10}$ kg s$^{-1}$ for the contour lines. The left panels present results from experiments with 3000 ppmv CO$_2$, while the right panels show results from experiments with 200,000 ppmv CO$_2$. The first and third rows correspond to experiments using CESM with 635 Ma paleogeography, and the second and fourth rows represent experiments using CESM with the modern Earth’s land configuration.}
\label{omega}
\end{figure}

The simulated results of hothouse climates are presented in Figure \ref{Figure_200000ppmv}, and the moderate climate simulations are shown in Figure \ref{Figure_3000ppmv} to provide a direct comparison between the climatologies of hothouse and moderate climates.

The most pronounced phenomenon in the hothouse climate is that a permanent NAIV (Figures \ref{Figure_200000ppmv}(a)\textendash(d)) occurs in the lower atmosphere as climate enters the hothouse state (Figures \ref{Figure_200000ppmv}(e)-(h)). NAIV is visible as the warmest air located within closed contours near $\sim$1000 hPa, centered in the subtropics. Analytically, the occurrence of NAIV can be defined as $T_{a}^{max} > T_{a}^{ns}$, where $T_{a}^{max}$ is the temperature of the warmest air in the vertical direction and $T_{a}^{ns}$ is the air temperature adjacent to the surface. NAIV occurs consistently across various models and land configurations.

NAIV mainly emerges over the subtropical ocean regions in the hothouse climate. Land configuration affects the distribution of NAIV. In the aquaplanet simulation, the atmospheric inversion only occurs in the subtropics between 15-30$^\circ$ (Figure \ref{Figure_200000ppmv}(l)). In simulations involving geography, the inversion predominantly manifests over subtropical ocean regions between 15-30$^\circ$ (Figures \ref{Figure_200000ppmv}(i)-(k)) and the strength can reach up to 4 K. Some weak inversion can be observed at various latitudes above the ocean, both higher and lower, but it is most pronounced in the subtropics. 

In hothouse climates, NAIV is stronger in the east ocean boundary than in the west when land is included. As seen in Figures \ref{Figure_200000ppmv}(i)-(k) and compared between Figures \ref{Figure_profile}(a) and (b), inversion is stronger in the east ocean boundary than in the west, regardless of models and land configurations. No NAIV is observed above land (Figures \ref{Figure_200000ppmv}(i)-(k)). In the aquaplanet simulation without ocean-land contrast, the strength of NAIV is zonally uniform (Figure \ref{Figure_200000ppmv}(l)).  

Another distinct phenomenon in hothouse climates is that the ocean surface is always cooler than the atmosphere above it (Figures \ref{Figure_200000ppmv}(m)-(p)). We define this phenomenon as ``surface inversion (SIV)", occurring when $T_{a}^{ns} > T_s$, where $T_s$ is the surface temperature. This phenomenon is also shown in Figures \ref{Figure_profile}(a)-(c), in which the surface (blue dots) is cooler than the near-surface air. Compared with other latitudes, SIV is strongest in the subtropics. Regionally, SIV is strongest in the coastal oceans. For example, a strong SIV can be seen near 120$^\circ$W, 40$^\circ$S in Figure \ref{Figure_200000ppmv}(m). Besides ocean regions, SIV can also be seen in the rainforests of tropical Africa and South America (Figure \ref{Figure_200000ppmv}(o)).

In addition, the temperature contrast between the ocean and land surfaces is significant in hothouse climates. As shown in Figures \ref{Figure_200000ppmv}(e)-(g), the land surface can be up to $\sim$30 K hotter than the ocean surface at the same latitude. Our simulations indicate that the average land surface temperature is 7 to 15 K higher than the average ocean surface temperature.

The substantial land-sea surface temperature contrast can trigger ascending air motion over some of the land areas. In moderate climates, there is large-scale descending motion over the continents near 30$^\circ$S in the 635 Ma paleogeography simulation (Figure \ref{omega}(a)) and over central Africa (near 30$^\circ$S) in simulations with modern Earth's land configuration (Figure \ref{omega}(c)), which is associated with the descending branch of the Hadley circulation (Figures \ref{omega}(e) and (g)). However, these descending areas transition to ascending areas in hothouse climates (Figures \ref{omega}(b) and (d)) by the large surface temperature contrast, which is significant enough to counteract the descending branch of the Hadley circulation.

\subsection{Mechanisms of Surface Inversion and Near-surface Atmospheric Inversion Formation}\label{mechanism}

The formation of SIV in hothouse climates can be explained using the surface energy budget equation following \cite{o2008hydrological} and \cite{wordsworth2013water}:
\begin{equation}
 c_p\rho_aC_D|\vec{v_a}|(T_{s}-T_{a}^{ns}) + \sigma[T_s^4-(T_a^{ns})^4]=F^{SW}_{abs} - F_L -F_{oc}.
 \end{equation}
Here, $F^{SW}_{abs}$ is the net shortwave radiation at the surface, $F_L$ is the surface latent cooling flux associated with evaporation, $F_{oc}$ is the energy loss through ocean heat transport, $c_p$, $\rho_a$, and $\vec{v_a}$ are the atmospheric specific heat at constant pressure, density, and velocity near the surface, $C_D$ is the drag coefficient, and $\sigma$ is the Stefan-Boltzmann constant. The first and second terms on the left-hand side are surface sensible heat and the net longwave emission at the surface. 

\begin{figure}
\noindent\includegraphics[width=\textwidth]{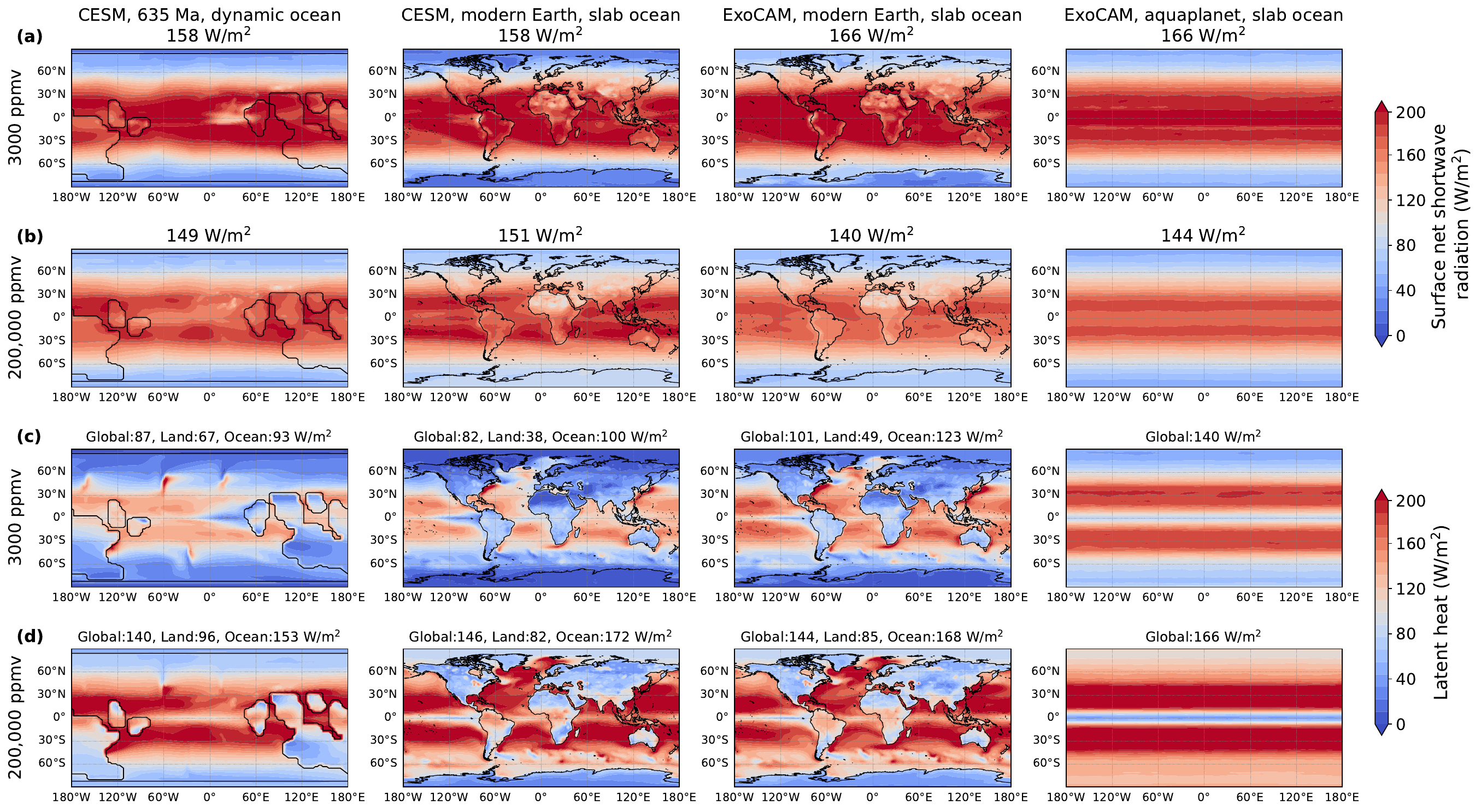}
\caption{The horizontal distribution of surface net shortwave radiative flux (a and b) and surface latent heat flux (c and d). 
Columns from left to right show results from experiments conducted with CESM 635 Ma paleogeography, CESM modern Earth's land configuration, ExoCAM modern Earth's land configuration, and ExoCAM aquaplanet. Rows (a) and (c) show results with 3000 ppmv CO$_2$, and rows (b) and (d) show results with 200,000 ppmv CO$_2$.
\label{Figure_latent}}
\end{figure}

\begin{figure*}[t!]
\centering
\includegraphics[width=0.95\textwidth]{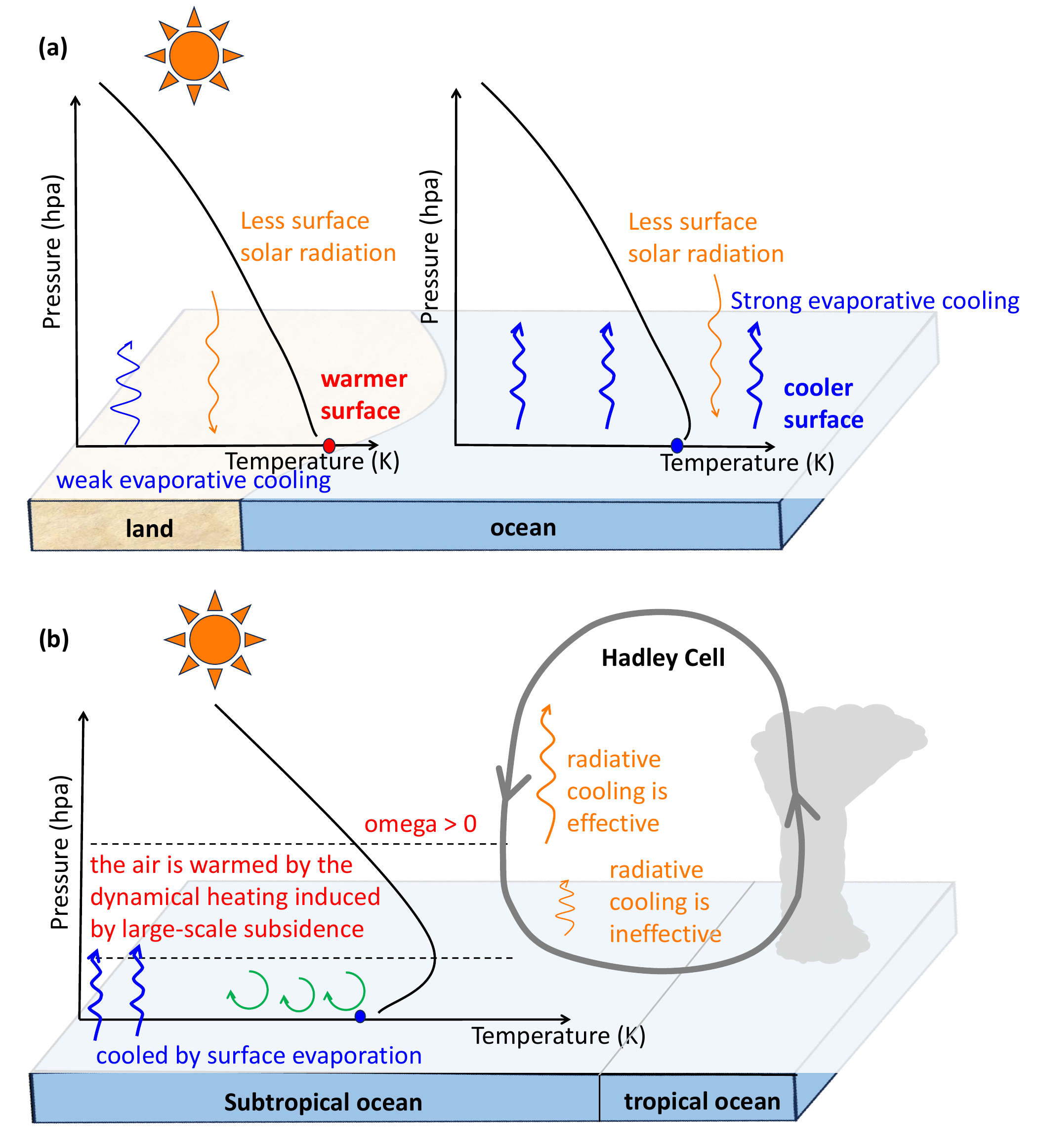}
\caption{Schematic diagrams of the formation of SIV (a) and NAIV (b). SIV is caused by strong surface evaporation and large atmospheric shortwave absorption. NAIV is caused by the dynamic heating induced by large-scale subsidence, lower-tropospheric radiative heating, and strong surface evaporation.
\label{Schemetic}}
\end{figure*}

The formation of SIV is due to the strong cooling associated with evaporation at the moist surface and the increased shortwave absorption by water vapor. In hothouse climates, the evaporative cooling is strong over the ocean (Figure \ref{Figure_latent}(d)). In addition, the water vapor content increases, leading to a larger atmospheric shortwave absorption \citep{wordsworth2013water} and a smaller $F^{SW}_{abs}$ (Figure \ref{Figure_latent}(b)). 
In experiments with a slab ocean, the ocean heat transport is prescribed and fixed. In experiments with a dynamic ocean, the change in ocean heat transport is small compared with surface latent heat flux and surface net shortwave radiation (Figures \ref{Figure_ocean_heat}(a) and (e)). Thus, a large $F_L$ and a small $F^{SW}_{abs}$ result in $F^{SW}_{abs}-F_L-F_{oc}<0$, leading to SIV formation ($T_{a}^{ns}$$ > T_{s}$) and resulting in negative surface sensible heat flux and longwave radiation directed downward \citep{o2008hydrological}. As shown in Figures \ref{Figure_200000ppmv}(m)-(p), SIV appears over all the ocean regions where there is large surface evaporation (Figure \ref{Figure_latent}(d)). It also emerges in tropical Africa and South America because surface evaporative cooling is also large in rainforests (Figure \ref{Figure_latent}(d)). It is worth noting that surface latent cooling flux does not always increase with surface warming. It increases from cold climates to $\sim$330 K and then decreases with warming after entering hothouse climates \citep{doi:10.1126/sciadv.ado2515}. However, the atmospheric shortwave absorption increases continuously, sustaining SIV. The above mechanisms are illustrated in the schematic diagram presented in Figure \ref{Schemetic}(a).

SIV is most pronounced at the ocean-land boundary due to the flow of warm air from the land, intensifying the temperature contrast between the ocean surface and the overlying air. 
For instance, the near-surface winds are westerly within 30$^{\circ}$S to 60$^{\circ}$S (Figures \ref{omega}(b) and (d)), directing the hot air towards the cooler ocean area, thereby amplifying SIV. Additionally, SIV is particularly strong over subtropical oceans primarily due to the substantial latent cooling in these regions (Figure \ref{Figure_latent}(d)).

The significant land-ocean surface temperature contrast observed in hothouse climates is primarily due to the substantial disparity in surface evaporation between the two. As mentioned above, in hothouse climates, the hot surface cannot be cooled effectively by emitting thermal infrared radiation into the space due to the closing of the water vapor absorption window. This makes surface evaporation the only mechanism to cool the surface. In moderate climates, experiments conducted with CESM 635 Ma paleogeography, CESM modern Earth's configuration, and ExoCAM modern Earth's configuration show differences in surface latent heat flux between the ocean and land of 26, 62, and 74 W m$^{-2}$, respectively (Figure \ref{Figure_latent}(c), positive associated with surface evaporation). In hothouse climates, these differences increase to 57, 90, and 83 W m$^{-2}$, respectively (Figure \ref{Figure_latent}(d)). The substantial contrast in surface latent cooling contributes significantly to the pronounced surface temperature difference between land and ocean.

The formation of NAIV can be explained using the atmospheric temperature tendency equation \citep{knutson1995time}:
\begin{equation}
    \frac{\partial T}{\partial t} = Q_\mathrm{dyn}  + Q_\mathrm{rad} + Q_\mathrm{conv} + Q_\mathrm{vdiff}.
\end{equation}
Here, $Q_\mathrm{rad}$ is the radiative heating rate, $Q_\mathrm{conv}$ is the convective heating rate, and $Q_\mathrm{vdiff}$ is the temperature tendency due to vertical diffusion. $Q_\mathrm{dyn}$ is the temperature tendency excluding radiation, convection, and vertical diffusion. It involves total dynamic heating ($-\vec{V} \cdot \nabla_h T - \omega \frac{\partial T}{\partial p}$, here $\vec{V}$ and $\omega$ are the horizontal and vertical velocity and $p$ is the air pressure), horizontal diffusion, and other residual terms such as the heating caused by filters and damping which are needed to keep the GCM running stably. 

The domain-averaged temperature tendencies are shown in Figures \ref{Figure_profile}(i)-(l). As all experiments have reached equilibrium, the net temperature tendency is close to zero. It is difficult to identify which processes are crucial in shaping the vertical temperature profile by checking their temperature tendency profiles in the equilibrium state above a single point. 
However, by comparing them in different regions, we can discern which processes contribute to the variations.

First, dynamic heating induced by large-scale subsidence is essential in forming NAIV. Compared between Figures \ref{Figure_profile}(i) and (k), the key distinction between ocean regions with and without NAIV is the dynamic heating rate. In regions with NAIV, the dynamic heating rate is positive, while in regions without inversion, it is negative above $\sim$1000 hPa. The radiative heating rate is similar, characterized by radiative cooling above $\sim$1000 hPa and heating below. The temperature tendency caused by vertical diffusion is also similar. 
Convective heating in these regions differs to maintain equilibrium. However, as convection typically serves as an adjustment to the temperature profile when convective instability occurs, we do not consider it as the primary process driving NAIV. The positive dynamic heating in the subtropical ocean region between 15-30$^\circ$ is caused by the adiabatic warming associated with the subsidence branch of the Hadley circulation (Figure \ref{omega}(h)). Conversely, the negative dynamic heating in the tropical ocean region results from the adiabatic cooling in the ascending branch of the Hadley circulation. The dynamic heating caused by large-scale subsidence warms the air near 1000 hPa to $\sim$335 K, warmer than the air below ($\sim$331 K), forming NAIV. While the air in the ascending region without this dynamic heating is still cooler than the air below, with no NAIV developing. Large-scale subsidence induced by the descending branch of the Hadley cell also helps form the eastern subtropical inversion in the present Earth's climate, particularly documented as trade-wind inversion \citep{schubert1995dynamical,johnson1999trimodal,carrillo2016characterization,https://doi.org/10.1002/joc.7151}. While without the combined effects of other exclusive phenomena in hothouse climates (discussed later), these inversions only occur over a small portion of the eastern subtropical ocean in the present climate and are absent in aquaplanet simulations (Figure \ref{Figure_3000ppmv}(l)).

Next, we find that LTRH is not independently sufficient to form NAIV. LTRH occurs in all the selected regions (Figures \ref{Figure_profile}(i)-(l)), whereas NAIV only develops over the subtropical ocean. 
Nonetheless, LTRH remains necessary. In moderate climates, the atmosphere in the subsidence branch of the Hadley circulation also undergoes dynamic heating, but this only leads to relatively weaker inversions (within $\sim$2 K), and these inversions are limited to very small portions of the eastern subtropical ocean in the present Earth's climate or no inversion in aquaplanet simulations, as previously mentioned. This is because radiative cooling and horizontal energy transport are effective in balancing the warming caused by subsidence. Once an inversion occurs, it can be quickly eliminated by radiative cooling. However, in hothouse climates, radiative cooling is not effective enough to cool the warmer air, so the inversion cannot be eliminated and can extend across the entire subtropical ocean. Additionally, with the effect of LTRH, the strength of NAIV can increase to as much as 4 K in hothouse climates (Figures \ref{Figure_200000ppmv}(i)-(k)). These make subtropical NAIV in hothouse climates distinct from the inversions observed over the eastern tropical ocean in the present Earth's climate.

Surface evaporative cooling also plays a role in forming NAIV. The relatively cool ocean surface caused by strong evaporation initiates the cooling of the air immediately above it through eddies and turbulence within the planetary boundary layer. Furthermore, this cooling effect extends to higher altitudes through vertical diffusion induced by the cool air. As observed in Figures \ref{Figure_profile}(i) and (j), the temperature tendency resulting from vertical diffusion exhibits a positive trend at the lowest level and turns negative between the lowest level and $\sim$1000 hPa, indicating downward energy transport triggered by the relatively cool ocean surface. This cooling enhances the temperature inversion around 1000 hPa, which makes NAIV formation different from that in the present climate where no planetary-scale surface inversion develops over the ocean (Figures \ref{Figure_3000ppmv}(m)-(p)).
It is important to clarify that vertical diffusion transports heat downgradient, meaning it cannot actively produce an inversion. Instead, it passively responds to the temperature profile, redistributing heat without directly contributing to the formation of NAIV. Additionally, surface evaporative cooling also increases atmospheric stability over the tropical oceans. As seen in Figures \ref{Figure_profile}(c) and (k), although no inversion develops, vertical diffusion cools the air below 1000 hPa, reducing the temperature lapse rate and thus enhancing atmospheric stability. 

In summary, the dynamic heating caused by large-scale subsidence plays a crucial role in the formation of NAIV. While LTRH is necessary, it is not sufficient independently to induce NAIV. Additionally, surface evaporation contributes to the development of NAIV. The underlying mechanisms are depicted in the schematic diagram shown in Figure \ref{Schemetic}(b).

The strength of NAIV in hothouse climates can also be explained by comparing the temperature tendencies between different regions. The stronger intensity of NAIV in the east ocean boundary, as opposed to the west, is attributed to the stronger subsidence observed in the eastern region (Figures \ref{Figure_profile}(e) and (f)). The stronger subsidence results in a larger dynamic heating rate, as indicated by the red lines in Figures \ref{Figure_profile}(i) and (j). This comparison provides additional evidence for the crucial role of large-scale subsidence in NAIV formation.
The absence of NAIV over land can be attributed to two factors. Firstly, the substantial temperature contrast between land and ocean surfaces induces upward motion over land, as discussed above. Secondly, the limited evaporative cooling over land results in a hot land surface. 

\begin{figure*}
\centering
\includegraphics[width=\textwidth]{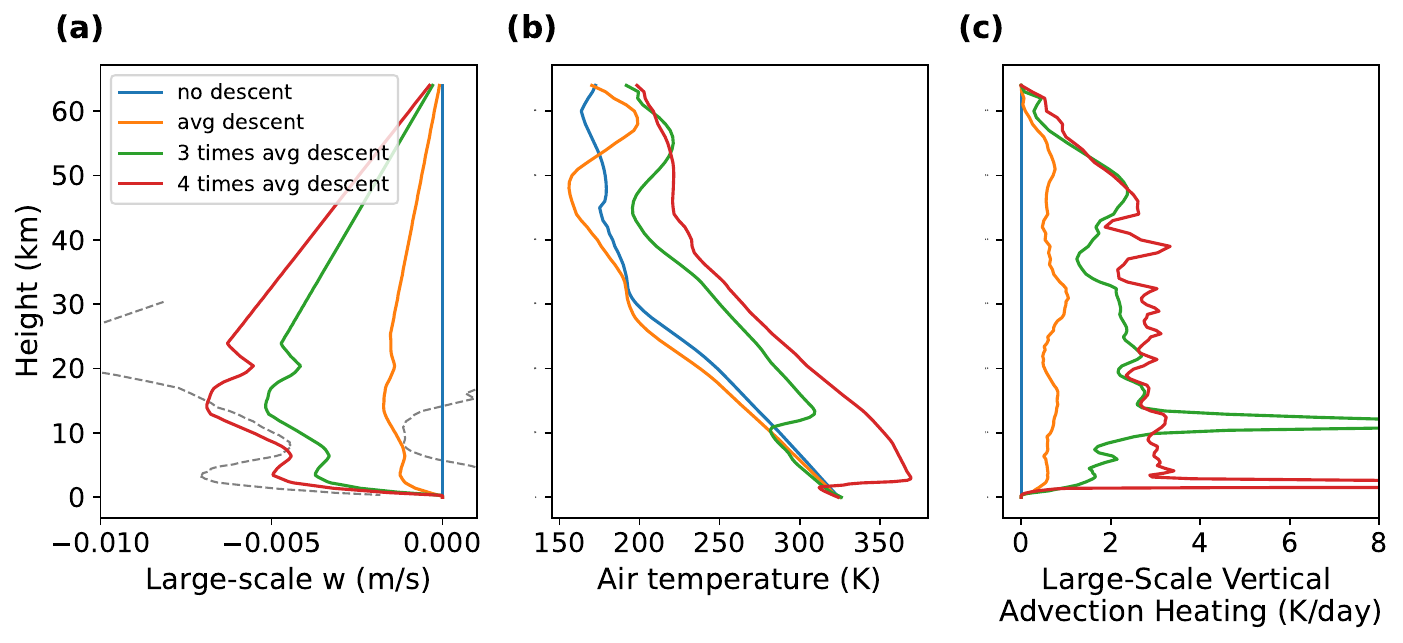}
\caption{Simulation results from cloud-resolving model SAM. Panel (a) shows the prescribed vertical velocity obtained from the vertical velocity of the descent area in CESM  (Figure \ref{Figure_profile}(e)) and then interpolated to SAM's vertical coordinate (orange line). Green and red lines represent three and four times the strength of the interpolated vertical velocity. The grey dashed lines represent the 10th and 90th percentiles of the monthly outputs from the CESM simulation over the past 10 years, with zero values excluded. Panels (b) and (c) show vertical profiles of air temperature and temperature tendency induced by the prescribed large-scale subsidence, respectively.
\label{CRM}}
\end{figure*}

\begin{figure}
\noindent\includegraphics[width=\textwidth]{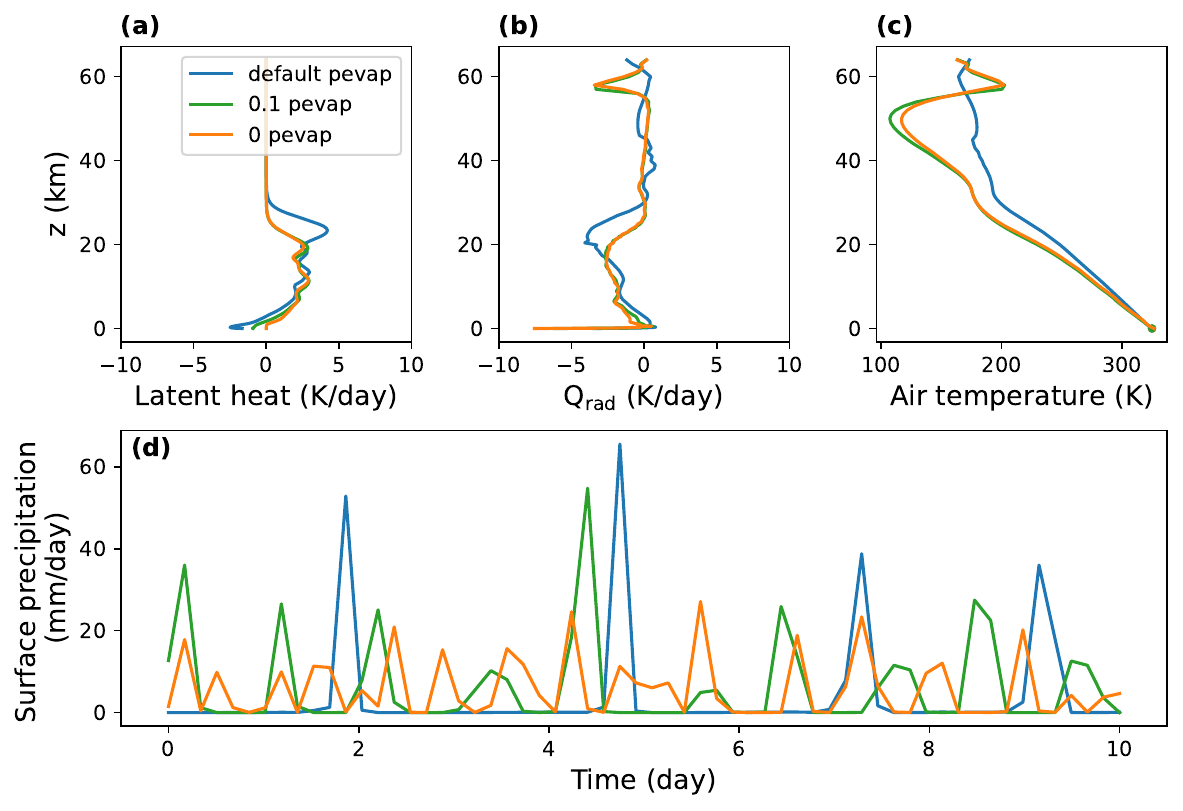}
\caption{Vertical profiles of latent heating (a), radiative heating (b), and air temperature (c) from the sensitivity tests for precipitation evaporation in SAM. Colored dots in panels (c) show the surface temperature. (d) Time series of surface precipitation.
The simulations are conducted with default coefficients of precipitation evaporation (blue), 0.1 times the default coefficients (green), and no precipitation evaporation (orange). 
\label{Reevaporation}}
\end{figure}

To further verify the importance of large-scale subsidence and explain the cloud-resolving simulation results in \citet{seeley2021episodic}, within which LTRH develops but no near-surface inversion is observed in hothouse climates, we manually add a prescribed large-scale subsidence in our small-domain SAM simulations. The prescribed large-scale subsidence is obtained from the regional mean vertical velocity (Figure \ref{Figure_profile}(e)) in the descending region over the ocean simulated by CESM (encircled by red rectangle in Figure \ref{Figure_200000ppmv}(j)) and then interpolated to the vertical coordinate of SAM (orange line in Figure \ref{CRM}(a)). Three different strengths of large-scale subsidence are used in the simulations (Figure \ref{CRM}(a)). 

In our simulation without prescribed large-scale subsidence, LTRH develops with the absence of NAIV (blue lines in Figures \ref{Reevaporation}(b) and (c)), consistent with previous simulated hothouse climates using small-domain CRMs \citep{seeley2021episodic,liu2023convection}. The near-surface inversion emerges when the large-scale subsidence and its induced heating are sufficiently strong (red and green lines in Figures \ref{CRM}(b) and (c)). This result further supports the importance of large-scale subsidence in forming NAIV. 

NAIV formation in SAM requires stronger subsidence than that shown in Figure \ref{Figure_profile}(e). This is because the vertical velocity shown in Figure \ref{Figure_profile}(e) is the 10-year- and domain-mean result, and more localized or intense subsidence events may be necessary to form NAIV. Additionally, the four times interpolated average vertical velocity falls within the model's original output range (Figure \ref{CRM}(a)), suggesting that the prescribed large-scale subsidence is reasonable for cloud-resolving modeling.

\citet{seeley2021episodic} found that a strong NAIV emerged when they turned off the evaporation of precipitation in their hothouse cloud-resolving simulations using DAM (Das Atmosphärische Modell). Thus, they argued that NAIV in hothouse climates simulated with GCMs may be a model's bias due to the highly idealized parameterization of this process.  To verify the impact of precipitation evaporation, we performed several sensitivity tests by changing the coefficient of precipitation evaporation in SAM, excluding large-scale subsidence. The results are shown in Figure \ref{Reevaporation}. We find that after reducing or turning off precipitation evaporation, no NAIV develops (Figure \ref{Reevaporation}(c)), contradictory to the results in \citet{seeley2021episodic}. These contradictory results between models may be caused by different parameterizations in precipitation evaporation or other microphysical processes. A more careful investigation of these discrepancies is needed in the future. Moreover, we observe that as we decrease precipitation evaporation in the simulations, the episodic deluge, which is characterized by short, intense bursts of precipitation separated by dry phases becomes less pronounced, with relatively shorter durations for the dry phases and less intense precipitation in the moist phases (Figure \ref{Reevaporation}(d)).

\subsection{A Conceptual Model}\label{1D model}

To illustrate the mechanisms forming NAIV discussed above, we develop a 1D conceptual model based on a two-stream double-gray atmosphere model \citep{pierrehumbert2010principles}. This model allows gray absorption in both shortwave and longwave, but the shortwave absorption coefficient $k_v$ is smaller than the longwave one $k_\mathrm{IR}$. The pure radiative equilibrium state of the double-gray atmosphere can be solved analytically \citep{pierrehumbert2010principles}, which suggests neither SIV nor NAIV formation and again implies that the radiative process alone is not sufficient to explain the formation of SIV and NAIV.

The 1D model considers only one atmospheric column, without incorporating horizontal transport. We include a non-radiative heating term $Q$ (with unit W\,m$^{-2}$) in the model to represent the combined effects of all physical processes excluding radiation. The heating rate profiles induced by $Q$ resemble that of the descending (Figure \ref{Figure_profile}(i)) or ascending (Figure \ref{Figure_profile}(k)) regions over the ocean in GCM simulations, depending on whether we prescribe dynamic heating or cooling in $Q$. A detailed description of $Q$ is provided in the derivation below. 

\begin{figure*}
\includegraphics[width=\textwidth]{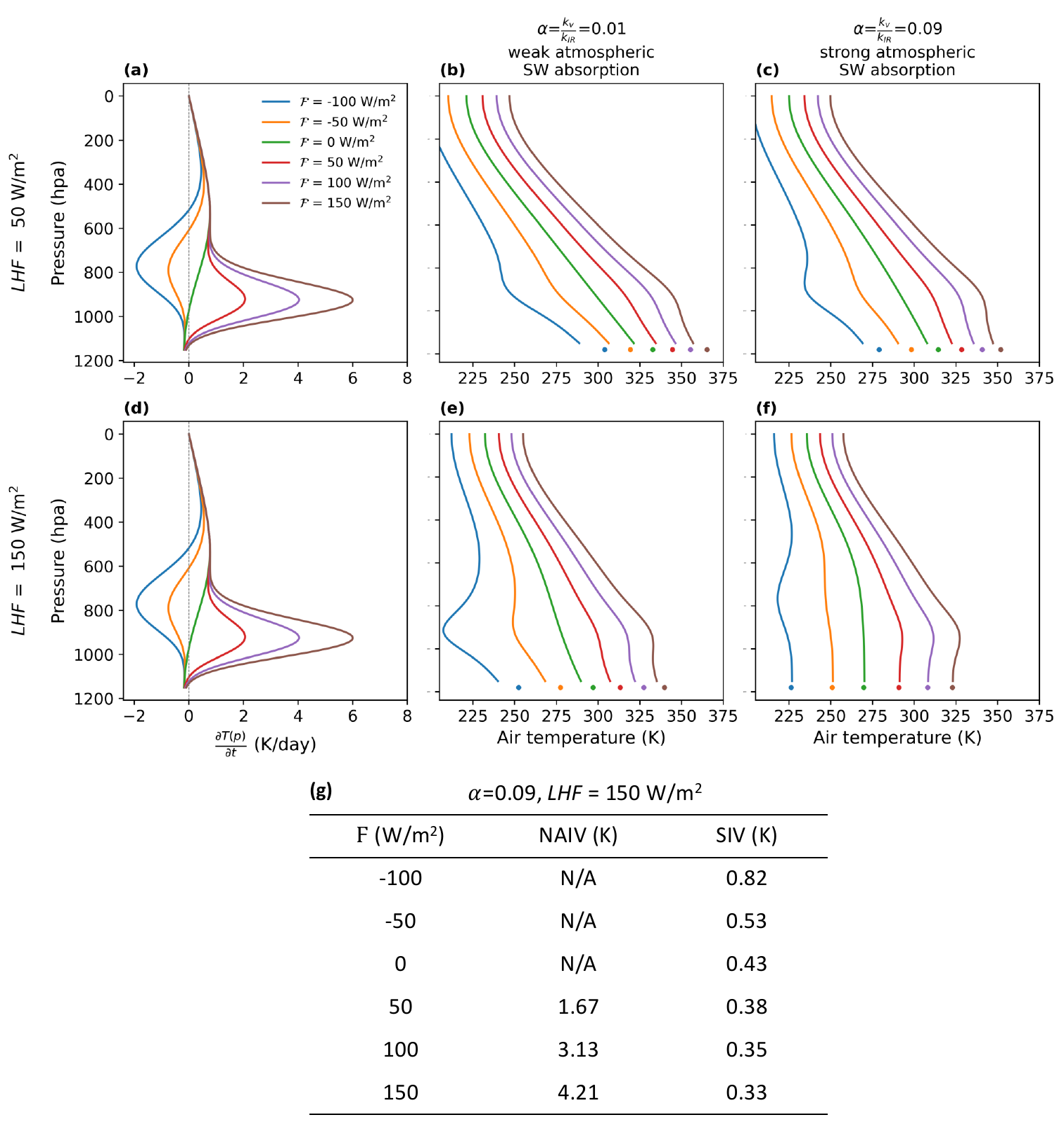}
\caption{Numerical results of the 1D conceptual model. Panels (a) and (d) show the combined heating rate $\frac{\partial T(p)}{\partial t}$ induced by $Q$. Panels (b), (c), (e), and (f) show the vertical profiles of air temperature (lines) and surface temperature (dots).
Results in the first and second rows are calculated with a small surface latent cooling of 50 W m$^{-2}$ and a large surface latent cooling of 150 W m$^{-2}$, respectively. The results in the second column are obtained assuming weaker atmospheric shortwave absorption, where the ratio of shortwave to longwave absorption coefficient $\alpha$ is set to 0.01. Results in the third column are calculated assuming a larger $\alpha$, 0.09. Color lines represent different external heating ($\mathcal{F}$) induced by large-scale subsidence (positive) or ascent (negative). Panel (g) shows the strength of NAIV and SIV in the simulations with 150 W m$^{-2}$, $\alpha=$0.09, and different $\mathcal{F}$.
\label{1D}}
\end{figure*}

We start the derivation from the two-stream radiative transfer equations of the double-gray atmosphere in the thermal infrared \cite[Chap.~4.3.4]{pierrehumbert2010principles}:
\begin{equation}
    \frac{d[F(\tau)^+-F(\tau)^-]}{d\tau} = F(\tau)^+ + F(\tau)^--\sigma T(\tau)^4,
\label{eqn1}
\end{equation}
\begin{equation}
    \frac{d[F(\tau)^++F(\tau)^-]}{d\tau} = F(\tau)^+ - F(\tau)^-.
\label{eqn2}
\end{equation}
$F(\tau)^{+}$ and $F(\tau)^{-}$ are the upward and downward radiative fluxes in the thermal infrared, $\tau$ is the average optical depth in the two-stream approximation, $\sigma$ is the Stefan–Boltzmann constant, and $T(\tau)$ is the air temperature \citep{goody1995atmospheric}.
The energy balance in each layer requires that the infrared cooling rate be balanced by the prescribed non-radiative heating $Q$ and shortwave absorption:
\begin{equation}
    \frac{d[F(\tau)^{+}-F(\tau)^{-}]}{d\tau} + Q(\tau) + \alpha F^{*}e^{-\alpha \tau} = 0,
\label{eqn3}
\end{equation}
where $\alpha$ is the ratio of the atmospheric absorption coefficient in the solar wavelengths to that in the thermal wavelengths ($\alpha = k_v / k_\mathrm{IR}$) and $F^*$ is the solar radiation at the top of the atmosphere. Integrating Equation (\ref{eqn3}) from 0 to $\tau$ yields the following:
\begin{equation}
    [F(\tau)^+-F(\tau)^-]|^{\tau}_{0} + \int^\tau_0\alpha F^*e^{-\alpha\tau'}d\tau' + \int^\tau_0 Q(\tau')d\tau' = 0.
\label{eqn4}
\end{equation}
We assume the column-integrated dynamic heating rate caused by large-scale subsidence or ascent as an external heating $\mathcal{F}$, in which a positive value indicates large-scale subsidence while a negative value indicates ascent. Ignoring the reflection of sunlight in the atmosphere and at the surface, the energy balance at the top of the atmosphere is:
\begin{equation}
    F(0)^{+}-F(0)^{-}-F^{*} = \mathcal{F}.
\label{eqn5}
\end{equation}
Thus, we can yield:
\begin{equation}
    F(\tau)^{+}-F(\tau)^{-} - F^*e^{-\alpha\tau} + \int^\tau_0 Q(\tau')d\tau'  = \mathcal{F}.
\label{eqn6}
\end{equation}
Combining Equation (\ref{eqn1}),  Equation (\ref{eqn2}), and Equation (\ref{eqn3}), we can obtain:
\begin{equation}
    \alpha^2F^*e^{-\alpha\tau} - \frac{dQ(\tau)}{d\tau} = F(\tau)^+-F(\tau)^- - 2\sigma\frac{dT(\tau)^4}{d\tau}.
\label{eqn7}
\end{equation}
Then combining Equation (\ref{eqn7}) with Equation (\ref{eqn6}) yields:
\begin{equation}
2\sigma\frac{dT(\tau)^4}{d\tau}=(1-\alpha^2)e^{-\alpha\tau}F^* - \int^\tau_0 Q(\tau')d\tau' +  \frac{dQ(\tau)}{d\tau} + \mathcal{F} .
\label{eqn8}
\end{equation}
Integrating Equation (\ref{eqn8}) from 0 to $\tau$ yields:
\begin{equation}
    2\sigma T(\tau)^4 = 2\sigma T(0)^4 + \frac{\alpha^2-1}{\alpha}F^*(e^{-\alpha\tau}-1) + Q(\tau)-Q(0) + \tau \mathcal{F} -\int^\tau_0 d\tau' \int^{\tau'}_0 Q(\tau'')d\tau''  .
\label{eqn9}
\end{equation}
Combining Equation (\ref{eqn1}) with Equation (\ref{eqn3}), and considering the top boundary condition Equation (\ref{eqn5}) can yield:
\begin{equation}
    2\sigma T(0)^4 = (1+\alpha)F^* + Q(0) + \mathcal{F}. 
\label{eqn10}
\end{equation}
Finally, by plugging Equation (\ref{eqn10}) into Equation (\ref{eqn9}), we can yield the air temperature at optical depth $\tau$:
\begin{equation}
    \sigma T(\tau)^4 = \frac{F^*}{2\alpha}[(\alpha^2-1)e^{-\alpha\tau}+\alpha+1] + \frac{(\tau+1)}{2}\mathcal{F}+\frac{1}{2}Q(\tau)-\frac{1}{2}\int^\tau_0 d\tau' \int^{\tau'}_0 Q(\tau'')d\tau''.
\label{eqn11}
\end{equation}
The air temperature profiles derived from Equation (\ref{eqn11}) are shown as color lines in Figures \ref{1D}(b), (c), (e), and (f). 

Next, we start to derive the expression for surface temperature. The surface is cooled by latent heat flux ($LHF$) through evaporation, so the surface energy balance can be written as:
\begin{equation}
    \sigma T^4_s + LHF =  F(\tau_{max})^- + F^*e^{-\alpha \tau_{max}},
\label{eqn12}
\end{equation}
where $\tau_{max}$ is the total atmospheric optical thickness in the thermal infrared. Combining Equation (\ref{eqn1}), Equation (\ref{eqn3}), and Equation (\ref{eqn6}), we can derive the expression of $F(\tau)^-$:
\begin{equation}
    F(\tau)^- = \sigma T(\tau)^4 -\frac{(\alpha+1)}{2}F^*e^{-\alpha\tau}-\frac{1}{2}Q(\tau)+\frac{1}{2}\int_0^\tau Q(\tau')d\tau' -\frac{\mathcal{F}}{2}.
\label{eqn13}
\end{equation}
Combining Equation (\ref{eqn11}), Equation (\ref{eqn12}), and Equation (\ref{eqn13}), we can obtain the expression for surface temperature:
\begin{equation}
    \sigma T^4_s = \frac{F^*}{2\alpha}[(\alpha-1)e^{-\alpha\tau_{max}}+\alpha+1] + \frac{\tau_{max}}{2}\mathcal{F}+\frac{1}{2}\int^{\tau_{max}}_0 Q(\tau)d\tau -\frac{1}{2}\int^{\tau_{max}} _0 d\tau \int^{\tau} _0 Q(\tau')d\tau'  -LHF.
\label{eqn14}
\end{equation}
in which, the atmospheric energy constraint should satisfy:
\begin{equation}
    \int^{\tau_{max}} _0 Q(\tau) d\tau = LHF + \mathcal{F}.
\label{eqn15}
\end{equation}
The surface temperatures derived from Equation (\ref{eqn14}) are shown as the colored dots in Figure \ref{1D}(b), (c), (e), and (f).
In the above calculation, $F^*$ is set to 240 W\,m$^{-2}$, considering solar insolation in 635 Ma with 0.25 planetary albedo, $\tau_{max}$ is 5, and $\sigma$ is 5.67$\times$10$^{-8}$ J\,s$^{-1}$\,m$^{-2}$\,K$^{-4}$.

To present the heating rate more clearly in units of K\,s$^{-1}$, we transform the optical thickness coordinate to the pressure coordinate, assuming a linear pressure-broadening case: $\tau$=$\tau_{max}$$(\frac{p}{p_{s}})^2$, in which $p$ and $p_s$ are the air pressure and the surface pressure (1.15 bar). The temperature tendency induced by $Q$ can be expressed as:
\begin{equation}
    \frac{\partial T(p)}{\partial t} = \frac{g}{c_p} Q(\tau) \frac{\partial \tau}{\partial p} = \frac{g}{c_p} Q(\tau)\frac{2\tau_{max}p(\tau)}{ p_{s}^2}.
\label{eqn16}
\end{equation}
Here, g is surface gravity (9.8 m\,s$^{-2}$) and $c_p$ is the air specific heat (1000 J\,kg$^{-1}$\,K$^{-1}$) at constant pressure.  The heating rate profiles calculated from Equation (\ref{eqn16}) are shown as color lines in Figures \ref{1D}(a) and (d).

As mentioned above, we employ the non-radiative heating term $Q$ to allow $\frac{\partial T(p)}{\partial t}$ to resemble the heating rate profiles in GCM simulations. Below $\sim$1000 hPa, the air is cooled by $Q$, which represents the cooling effect caused by vertical diffusion driven by the cool ocean surface. We assume this cooling is always proportional to the strength of surface evaporation. Above 600 hPa, the air undergoes convective heating. Between about 800 to 1000 hPa, we apply dynamic heating to represent large-scale subsidence over the subtropical ocean and apply dynamic cooling to represent large-scale ascent over the tropical ocean. Mathematically, $Q$ is in the form of a superposition of three Gaussian functions:
\begin{equation}
    Q(\tau)=\frac{\mathcal{F}}{\sqrt{2\pi}\sigma_1}exp[-\frac{(\tau-0.64\tau_{max})^2}{2\sigma_1^2}]-\frac{\beta LHF}{\sqrt{2\pi}\sigma_2}exp[-\frac{(\tau-\tau_{max})^2}{2\sigma_2^2}]+\frac{(\beta+2) LHF}{\sqrt{2\pi}\sigma_3}exp(-\frac{\tau^2}{2\sigma_3^2}),
\label{eqn17}
\end{equation}
or
\begin{equation}
    Q(\tau)=\frac{\mathcal{F}}{\sqrt{2\pi}\sigma_1}exp[-\frac{(\tau-0.4\tau_{max})^2}{2\sigma_1^2}]-\frac{\beta LHF}{\sqrt{2\pi}\sigma_2}exp[-\frac{(\tau-\tau_{max})^2}{2\sigma_2^2}]+\frac{(\beta+2) LHF}{\sqrt{2\pi}\sigma_3}exp(-\frac{\tau^2}{2\sigma_3^2}),
\label{eqn18}
\end{equation}
when $\mathcal{F}$ is positive or negative, respectively.
The three terms on the right-hand side of Equations (\ref{eqn17}) and (\ref{eqn18}) represent adiabatic heating or cooling induced by subsidence or ascent, cooling caused by the cool surface, and convective heating rate. When $\mathcal{F}$ is positive, $\sigma_1$, $\sigma_2$, and $\sigma_3$ are set to 0.6, 1.6, and 1.5 respectively, while when 
$\mathcal{F}$ is negative, $\sigma_1$, $\sigma_2$, and $\sigma_3$ are set to 0.8, 1.6, and 1.5. $\beta$ is 0.2. 

The main findings obtained from the conceptual model can be summarized as follows:
\begin{enumerate}
   \item As shown in Figures \ref{1D}(f) and (g), larger dynamic heating can induce stronger NAIV. The strength of NAIV ($T_{a}^{max}-T_{a}^{ns}$) is  1.67, 3.13, 4.21 K assuming the external heating induced by subsidence $\mathcal{F}=$ 50, 100, and 150 W\,m$^{-2}$, respectively. No NAIV occurs without heating induced by large-scale subsidence.
   \item When comparing Figures \ref{1D}(c) and (f), and shown in Figure \ref{1D}(g), it is evident that strong surface evaporation can induce SIV, contributing to the overall formation of NAIV. With large surface evaporation, the surface temperature (circles) is cooler than the near-surface air (Figure \ref{1D}(f)) and both SIV and NAIV develop.
   \item Comparing Figures \ref{1D}(e) and (f), stronger atmospheric shortwave absorption contributes to the formation of SIV. With a fixed $Q$ profile, $\alpha$ (the ratio between shortwave and longwave absorption coefficient) affects the surface energy balance by determining how much solar radiation reaches the surface (Equation (\ref{eqn12})). Larger $\alpha$ signifies a larger atmospheric shortwave absorption and smaller absorbed shortwave radiation at the surface ($F_{abs}^{SW}$). 
\end{enumerate}

The conceptual model verifies the importance of large-scale subsidence and strong surface evaporation in forming NAIV and the crucial role of strong surface evaporation and atmospheric shortwave absorption in SIV formation. Since the radiative heating rate is determined by other non-radiative heating rates to sustain equilibrium in our model, the contribution of LTRH is not examined. 

\section{Conclusions and Discussion}\label{conclusions}
In this study, we investigate the underlying mechanisms of the formation of near-surface atmospheric inversion (NAIV) and surface inversion (SIV) in hothouse climates. 
We propose that the dynamic heating induced by large-scale subsidence is essential in forming NAIV. The lower-tropospheric radiative heating (LTRH) is necessary but not independently sufficient to induce NAIV. Furthermore, the surface evaporation also contributes to the formation of NAIV. While both NAIV and LTRH are well-recognized in hothouse climate models, the specific role of LTRH in NAIV formation has not been thoroughly explained in previous research. Our study combines the effects of LTRH, dynamic subsidence, and surface evaporation to provide a comprehensive explanation. SIV forms in hothouse climates because of the strong surface evaporation and large atmospheric shortwave absorption. The horizontal advection of warm air from the land area to the cooler ocean can amplify SIV over the coastal oceans. To further illustrate the mechanisms of NAIV and SIV formation, we develop a 1D conceptual model based on the two-stream double-gray atmosphere. This 1D model verifies the importance of large-scale subsidence and surface evaporation in forming NAIV, as well as the significance of strong surface evaporation in the development of SIV.

In previous studies, NAIV emerges in hothouse climate simulations employing GCMs \citep{wolf2015evolution, popp2016transition} but is not observed in small-domain CRM simulations \citep{seeley2021episodic,seeley2023moist}. We suggest this discrepancy is because large-scale subsidence and the associated adiabatic warming are not included in previous CRM simulations.
We find that with reasonably-prescribed large-scale subsidence, NAIV emerges. This result confirms our presumption, further supports the essential of large-scale subsidence in forming NAIV, and also bridges the gap between GCM and CRM simulations.

\citet{seeley2021episodic} found NAIV emerged in their CRM simulations after turning off the evaporation of precipitation and argued that NAIV in hothouse climates simulated with GCMs may be a model's bias due to the highly idealized parameterization of this process. However, when we reduce or turn off the re-evaporation process in SAM, no NAIV  develops. These contradictory results between models may be caused by different parameterizations in precipitation evaporation or other microphysical processes and need to be investigated in the future. NAIV is also observed in a 1D radiative-convective equilibrium model \citep{wordsworth2013water}, however, given that 1D models typically lack boundary layer schemes and incorporate various assumptions, their results may not fully represent the real atmosphere and should be reassessed using more complex CRM or GCM simulations. 

Subtropical NAIV is not exclusive to hothouse climates; it is also observed under Earth's current climate and has been analyzed in previous papers, particularly as the trade-wind inversion driven by large-scale subsidence in the descending branch of the Hadley cell and the upwelling of cold ocean water \citep{schubert1995dynamical,johnson1999trimodal,carrillo2016characterization,https://doi.org/10.1002/joc.7151}. However, subtropical NAIV in the present climate differs from that in hothouse climates in several ways. Firstly, in the current climate, subtropical NAIV is mostly limited to small regions over the eastern subtropical oceans, with a typical strength within $\sim$2 K \citep{carrillo2016characterization,https://doi.org/10.1002/joc.7151}, and does not appear in aquaplanet simulations (Figure \ref{Figure_3000ppmv}(l)). This is because radiative cooling and horizontal energy transport in today’s climate effectively counteract the adiabatic heating induced by large-scale subsidence. In contrast, in hothouse climates, NAIV formation is strongly influenced by LTRH, causing it to expand across the entire subtropical ocean and increase its strength to as much as 4 K (Figures \ref{Figure_200000ppmv}(i)-(j)). Secondly, NAIV formation in hothouse climates is also influenced by a planetary-scale cooler ocean surface, resulting from strong evaporative cooling, which is not a feature of the present-day climate. Lastly, with its greater strength and wider spatial extent, NAIV in hothouse climates has a larger impact on planetary climate compared to the smaller-scale inversions observed in the present-day subtropical eastern oceans.

Large-scale temperature inversions also occur in the boundary layer of the Arctic and Antarctic under the current climate, as shown in Figures \ref{Figure_3000ppmv}(i)-(k) and other atmospheric soundings and model simulations \citep{Zhang_Seidel_Golaz_Deser_Tomas_2011, Ruman_Monahan_Sushama_2022}.
Different from subtropical NAIV, these surface-based inversions mainly form due to surface radiative cooling, where surface longwave emission exceeds absorbed solar radiation and the advection of warm air above the surface \citep{Bradley_Keimig_Diaz_1992, Bradley_Keimig_Diaz_1993}. They are most frequent during winter, when the polar regions receive less solar radiation, and the larger meridional temperature gradient favors stronger advection of relatively warm air from lower latitudes. 

Our study primarily focuses on Earth or potentially habitable Earth-twin exoplanets (asynchronous rotators with atmospheric circulation patterns similar to Earth) orbiting Sun-like stars. However, when considering the likely more abundant Earth-like habitable planets around M dwarfs, it is important to note that M stars emit more near-infrared radiation than Sun-like stars. Since water vapor absorbs near-infrared radiation effectively, LTRH may occur at lower surface temperatures on exoplanets around M stars \citep{seeley2021episodic}, possibly leading to the formation of NAIV and SIV under these cooler conditions. Nonetheless, their formation mechanisms are expected to remain consistent with this study. For tidally-locked potentially habitable exoplanets, the atmospheric circulation is characterized by strong ascent at the sub-stellar regions and descent on the night side. Their radiation and circulation patterns may lead to the formation of an inversion layer within the night-side boundary layer, driven by radiative cooling of the surface and dynamic heating caused by large-scale subsidence \citep{joshi_simulations_1997,wordsworth_atmospheric_2015}. This mechanism is analogous to the formation of Earth's polar night boundary inversion \citep{joshi_earths_2020}.

\begin{figure}
\centering
\noindent\includegraphics[width=0.85\textwidth]{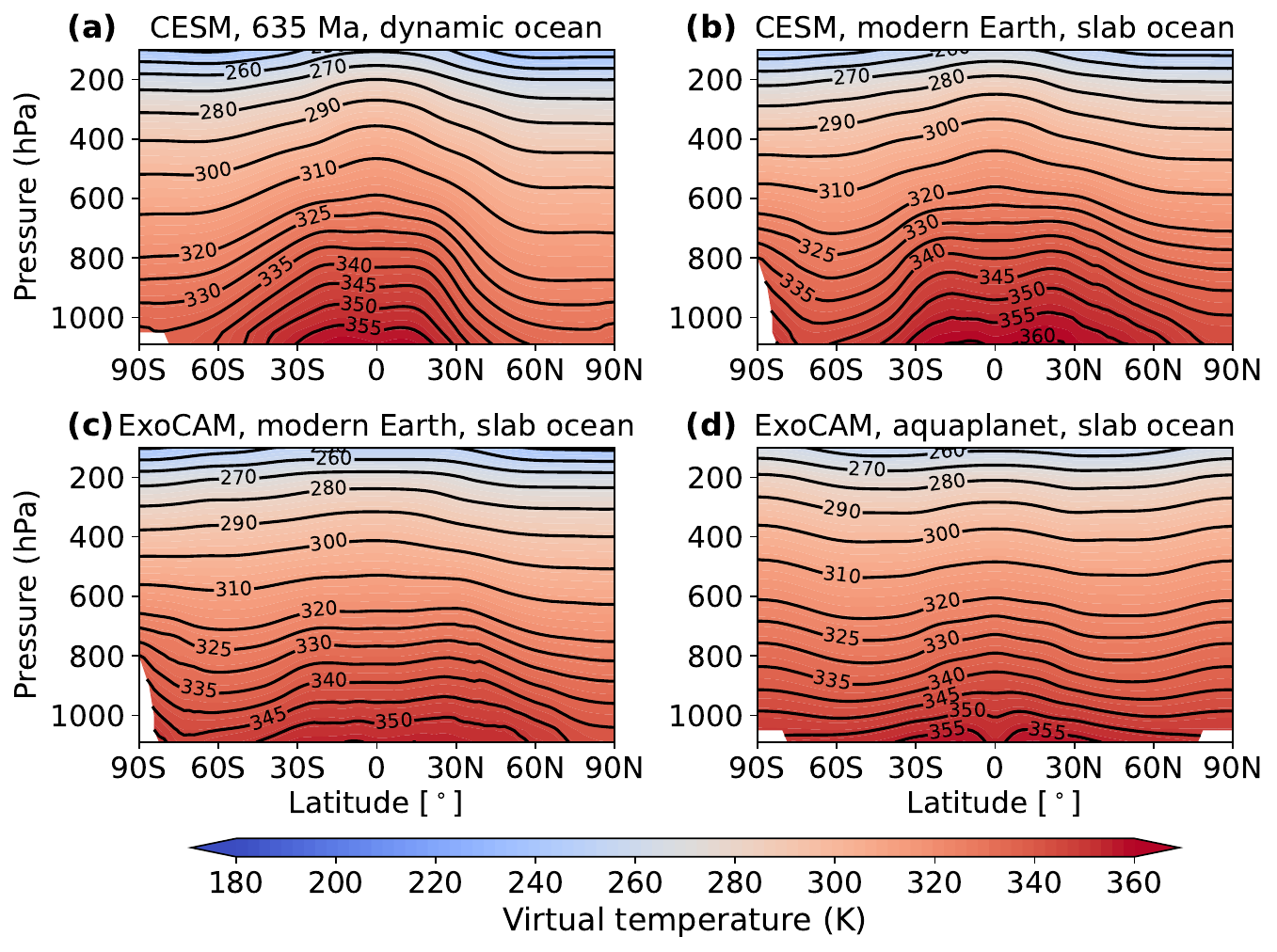}
\caption{Latitude versus pressure distribution of virtual temperature from GCM simulations with 200,000 ppmv CO$_2$. Panels (a), (b), (c), and (d) are results from experiments conducted with CESM 635 Ma paleogeography, CESM modern Earth’s land configuration, ExoCAM modern Earth’s land configuration, and ExoCAM aquaplanet, respectively.
\label{virtual_T}}
\end{figure}

The emergence of NAIV in hothouse climates indicates a stable atmospheric stratification, not just in subtropical regions, but also globally. This stable atmosphere can affect atmospheric circulation, convection, and the hydrological cycle. For example, NAIV may hinder deep convection rooted from the surface and thereby reduce precipitation \citep{doi:10.1126/sciadv.ado2515}. The reduced precipitation can decrease the weathering rate in hothouse climates \citep{walker1981negative}, so the recovery from the
post-snowball hot climate should be longer than that previously assumed \citep{hoffman1998neoproterozoic}. Note that, in addition to the temperature vertical gradient (thermal buoyancy), atmospheric stability is also influenced by vapor buoyancy \citep{yang2020incredible,seidel2020lightness,yang2022substantial}. While this paper primarily focuses on thermal buoyancy, vapor buoyancy is not negligible in the atmospheric stability in hothouse climates. As demonstrated in Figure \ref{virtual_T}, there is no inversion in virtual temperature in the hothouse climate GCM simulations, suggesting that vapor buoyancy plays an important role in enhancing atmospheric instability, compensating for the increased stability by thermal inversion. However, even when accounting for the effects of vapor buoyancy, near-surface atmospheric stability still increases in hothouse climates, as evidenced by a much smaller global-mean virtual temperature vertical gradient below 1000 hPa (not shown).

\begin{figure}
\noindent\includegraphics[width=\textwidth]{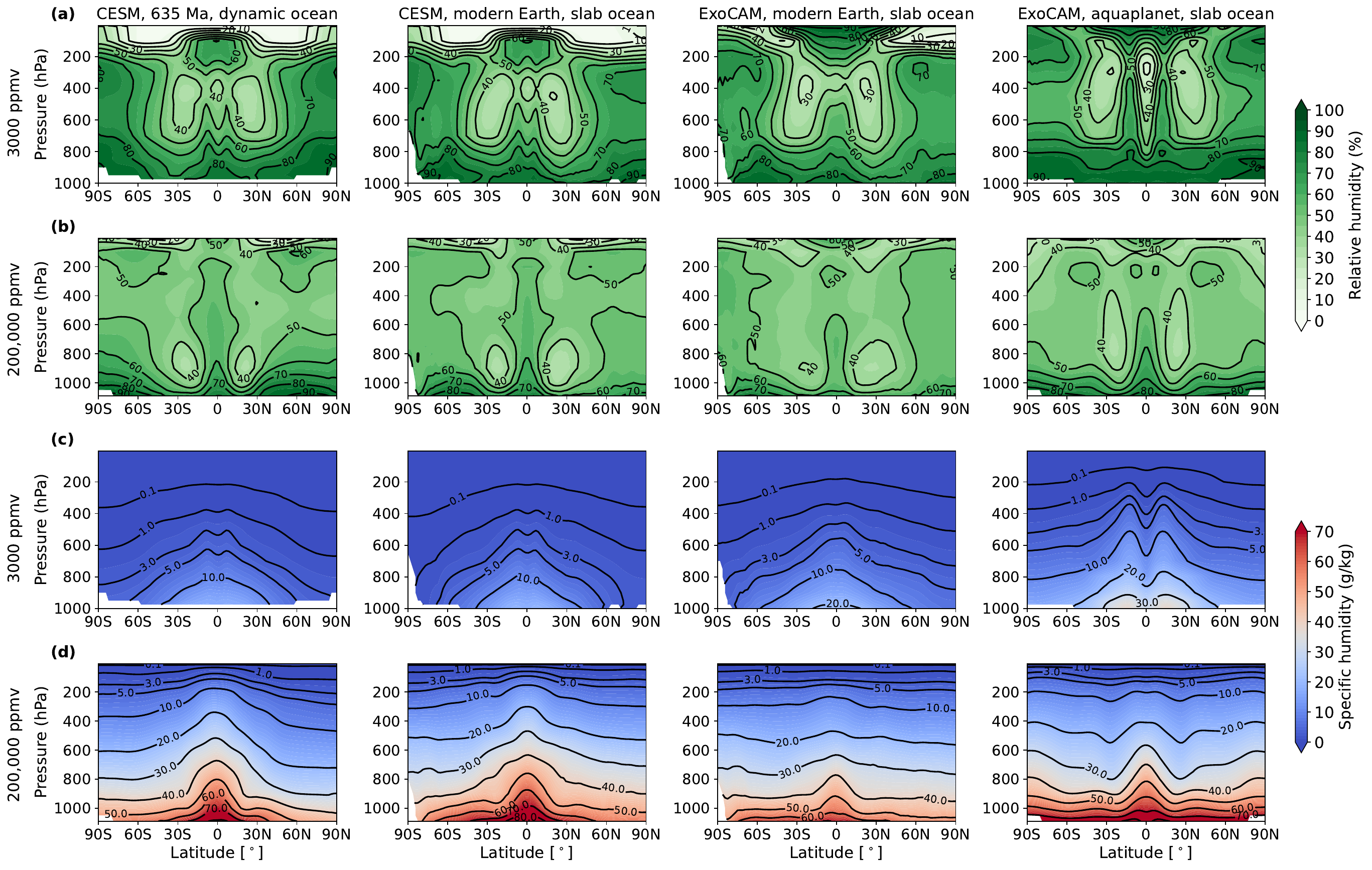}
\caption{The latitude versus pressure distribution of relative humidity (a and b) and specific humidity (c and d).  Columns from left to right show results from experiments conducted with CESM 635 Ma paleogeography, CESM modern Earth's land configuration, ExoCAM modern Earth's land configuration, and ExoCAM aquaplanet. Rows (a) and (c) show results with 3000 ppmv CO$_2$ and rows (b) and (d) show results with 200,000 ppmv CO$_2$.
\label{Figure_RH}}
\end{figure}

The stable atmosphere corresponding with NAIV reduces the relative humidity above (Figure \ref{Figure_RH}(b)). This leads to a lower water vapor mixing ratio in the stratosphere (Figure \ref{Figure_RH}(d)) compared to a hothouse climate condition without NAIV, thereby delaying the onset of the moist greenhouse state. As a result, surface water can remain on Earth for a longer timescale under a brightening Sun in the future or on potentially habitable exoplanets near the inner edge of the habitable zone.

\begin{acknowledgments}
We thank Mengyu Wei, Jiawenjing Lan, and Jian Zhang for their help in modifying the paleogeography in CESM. We thank Eric Wolf and the CESM team for releasing model ExoCAM and CESM. We are grateful to Jacob T. Seeley for his shared data. We also thank the thoughtful comments from an anonymous referee that improved the quality of the paper. 
Simulation data, as well as the codes used for analysis and figure plotting in this study, are stored and accessible at https://doi.org/10.5281/zenodo.15090054.
J.Y. was supported by NSFC under grant nos. 42441812 and 42275134. F.D. was supported by the Fundamental Research Funds for the Central Universities (Peking University).
\end{acknowledgments}

\bibliography{sample631}{}
\bibliographystyle{aasjournal}

\end{document}